\documentclass[aps,pre,twocolumn,superscriptaddress,showpacs,showkeys]{revtex4}
\usepackage{epsfig}
\usepackage{graphicx}
\usepackage{amsmath,color}
\usepackage[english]{babel}

\begin{document}

\title{Improving the performance of algorithms to find communities in networks}

\author{Richard K. Darst}
\affiliation{Department of Biomedical Engineering and Computational
  Science, Aalto University School of Science, P.O.  Box 12200,
  FI-00076, Finland}
\author{Zohar Nussinov}
\affiliation{Physics Department, Washington University in
  St. Louis, CB 1105,
Washington University,
One Brookings Drive,
St. Louis, MO 63130-4899, USA}
\author{Santo Fortunato}
\affiliation{Department of Biomedical Engineering and Computational
  Science, Aalto University School of Science, P.O.  Box 12200,
  FI-00076, Finland}

\begin{abstract}
Most algorithms to detect communities in networks typically work
without any information on the cluster structure to be found, as one
has no {\it a priori} knowledge of it, in general. Not surprisingly,
knowing some features of the unknown partition could help its
identification, yielding an improvement of the performance of the
method. Here we show that, if the number of clusters were known
beforehand, standard methods, like
modularity optimization, would considerably gain in accuracy, mitigating
the severe resolution bias that undermines the reliability of the
results of the original unconstrained version. The number of clusters
can be inferred from the spectra of the recently introduced
non-backtracking and flow matrices, even in benchmark graphs with
realistic community structure. The limit of such two-step procedure is
the overhead of the computation of the spectra.
\end{abstract}

\maketitle

\section{Introduction}
\label{intro}

Community structure is one of the most important features of complex
networks. Communities, or clusters, are subgraphs of a network with an
unusually high density of edges, with respect to the average
edge density of the network as a whole. Nodes in the same community
tend to share the same attributes and/or roles within the network, and
their identification might lead to the discovery of unknown features
purely from topological inputs. This is why the quest for methods to detect communities in
networks has become one of the hottest topics in network
science~\cite{girvan02,fortunato10}.

In general, the only preliminary information available to any
algorithm is the topology of the network, i.e.\ which pairs of nodes are
connected to each other and which are not. Anything about the
cluster structure to be found is unknown and is supposed to be given
as output of the procedure. It would be valuable to have some
information on the unknown partition of the network, as one could
reduce considerably the huge space of possible partitions to explore,
and increase the chance of successfully identifying the communities. 
In particular, if the number of clusters were known and could be given
as input/constraint, methods deliver more accurate partitions.

The problem is that it is difficult to have access to the number of clusters.
Fortunately, recent work has shown that it is possible to derive this
information from the spectra of the non-backtracking matrix~\cite{krzakala13} and the
flow matrix~\cite{newman13}, at least on the classic version of the
planted partition model~\cite{condon01}, where clusters have identical
size and nodes the same degree (on average). We show that the
prediction of the number of clusters remains accurate as well on the
LFR benchmark graph~\cite{lancichinetti08}, which extends the original planted partition
model by introducing realistic features of community structure, i.e.\
heterogeneous distributions of degrees and cluster sizes. The
prediction is more reliable the larger the graph size.

Therefore, we propose to improve the performance of community
detection algorithms via a two-step approach: first one infers the
number $q$ of clusters from the spectrum of the non-backtracking or
the flow matrix; then one uses this number as additional input of
the algorithm. We show that popular methods, like the optimization of
the modularity by Newman and Girvan~\cite{newman04c}, become much more
accurate this way. This is remarkable, as the direct optimization of
modularity is known to have a limited resolution, which may prevent the
method from identifying the correct scale of the communities, even
when the latter are very pronounced~\cite{fortunato07,good10}. Knowing the number of clusters $q$, and
constraining the optimization of the measure among partitions with
fixed $q$, one can mitigate the problem. 

The computation of the spectra is unfortunately quite heavy, and
represents the bottleneck of the procedure, making the analysis of
large networks ($\gtrsim 1\,000\,000$ links) basically unfeasible. However, our results
are encouraging and might stimulate the development of quicker
approximate heuristics than the ones currently available.

In Section II we reveal the importance of knowing the number of
clusters for the results of popular algorithms. Section III tests the reliability of the prediction of the
number of clusters in the standard planted partition model and in the
LFR benchmark graph. In Section IV we introduce a two-step approach
consisting of a fast greedy optimization of the modularity
by Newman and Girvan for partitions with a fixed number of clusters
and we show the superiority of its performance with respect to the
exhaustive modularity optimization via simulated annealing on the LFR benchmark.
Our findings are summarized in Section V.

\section{Constrained versus unconstrained community detection}
\label{sec2}

Stochastic block-models are the best known and most
exploited class of models of networks with community structure. A
graph with $N$ nodes 
is divided into $q$ groups of equal size $n=N/q$, and the probability of nodes $i$
and $j$ to be linked is given by $p_{rs}$, where $r$ and $s$ are the
groups of $i$ and $j$, respectively. The principle is that the linking
probability of any two nodes only depends on the memberships of the
nodes. Despite its simplicity, this model can generate a wide variety
of scenarios. Here we shall focus on the simplest one, where
$p_{rs}=p_{in}, \forall r=s$ and $p_{rs}=p_{out}, \forall r\neq
s$. Here, if $p_{in}>p_{out}$, the expected number of neighbors of a
node within its group exceeds the expected number of neighbors of the
node in each of the remaining $q-1$ groups, so the groups are
communities according to the general intuition.
This version coincides with the planted partition
model by Condon and Karp~\cite{condon01}, and has generated
popular benchmark graphs that are regularly used to test the
performance of community detection techniques, like the four-groups
test~\cite{girvan02} and the LFR benchmark~\cite{lancichinetti08}.

According to the model, clusters are present and should be detectable so
long as $p_{in}>p_{out}$. However,
recent works~\cite{reichardt08,decelle11,decelle11b,nadakuditi12,floretta13} have pointed out that, in the
limit of sparse graphs, i.e.\ of networks of infinite size but finite
average degree, random fluctuations make the detectability of clusters
possible only for $p_{in}>p_{out}+\Delta$, where
$\Delta$ is a function of the parameters $n, q, p_{in}, p_{out}$. So
it is necessary for the clusters to have an internal
edge density higher by some finite amount than the external one(s) to overcome
the noise coming from the inherent stochasticity of the network's
construction. By introducing the average internal and external degree
of a node in a cluster, $\mu_{in}=(n-1)p_{in}$
and $\mu_{out}=np_{out}$, respectively, the formula of the detectability
limit derived in Refs.~\cite{decelle11,nadakuditi12} reads
\begin{equation}
\mu_{in}-\mu_{out} = \sqrt{\mu_{in}+(q-1)\mu_{out}}\,\,.
\label{eq1}
\end{equation}
Typically, one fixes the value of the internal average degree
$\mu_{in}$ and obtains the limit value of $\mu_{out}$ from
Eq.~\ref{eq1}, beyond which no method can do better than a random
guess, which yields $1/q$, where $q$ is the number of clusters.

\begin{figure}[h]
\begin{center}
\includegraphics[width=\columnwidth]{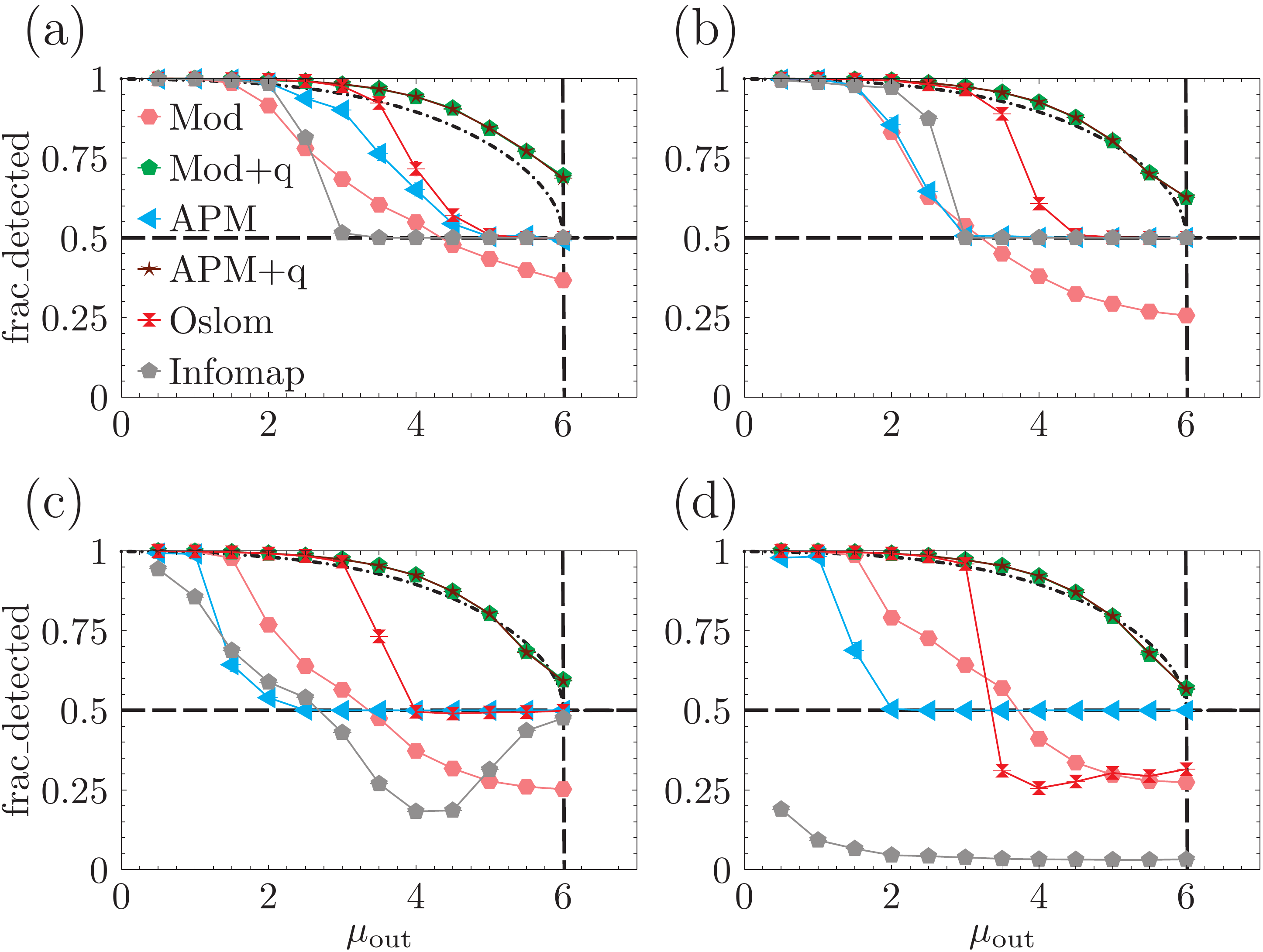}
\caption{(Color online) Fraction of correctly detected nodes for the
  planted partition model with $q=2$ clusters, average internal degree
  $\mu_{in}=10$ and four values for the size of the clusters: (a) $n=50$,
  (b) $n=200$, (c) $n=500$ and (d) $n=1000$. Symbols indicate
  the performance curves of different popular methods of community
  detection. For modularity optimization (Mod) and the Absolute Potts
  Model (APM) we show two curves, referring to the results of the
  method in the absence of any information on the number of clusters
  of the planted partition, and when such information is fed into the
  model as initial input. In both cases, knowing the number of
  clusters beforehand leads to a much better performance. Modularity's
 limit performance (dot-dashed curve), as estimated by Nadakuditi and
 Newman~\cite{nadakuditi12}, is attained only when the number of
 clusters is known. Otherwise modularity optimization performs rather
 poorly, as expected. The detectability limit is shown as a vertical
 line at $\mu_{out}=6$.}
\label{fig1}
\end{center}
\end{figure}

In Fig.~\ref{fig1} we show the performance of several popular community
detection algorithms on such
model, in the simple case of two clusters ($q=2$). Each panel corresponds to a different value of the cluster size
$n$, ranging from $50$ to $1000$. The average internal degree
$\mu_{in}$ is fixed to $10$. Given these parameters, Eq.~\ref{eq1} 
yields a value $\mu_{out}^{det}=6$ for the detectability limit (dashed
vertical line), regardless of the cluster size $n$, and we vary $\mu_{out}$
from $0$ to $\mu_{out}^{det}$.
The dot-dashed
line indicates the limit performance of spectral modularity
optimization, analytically derived in Ref.~\cite{nadakuditi12}. 
Performance is computed as the fraction of correctly detected
nodes~\cite{nadakuditi12}. If all nodes are correctly assigned to
their planted clusters, the fraction of correctly detected nodes is
$1$. As long as clusters are detectable, it stays above the value
corresponding to random assignment (here $1/2$, marked by the
horizontal dashed line).
The algorithms we adopted are: modularity optimization via simulated
annealing (Mod)~\cite{guimera04}, the Absolute Potts
Model (APS) \cite{ronhovde10}, OSLOM~\cite{lancichinetti11} and
Infomap~\cite{rosvall08}.
Further information on all methods is available in App.~\ref{sec:methods}.
\begin{figure}[h]
\begin{center}
\includegraphics[width=\columnwidth]{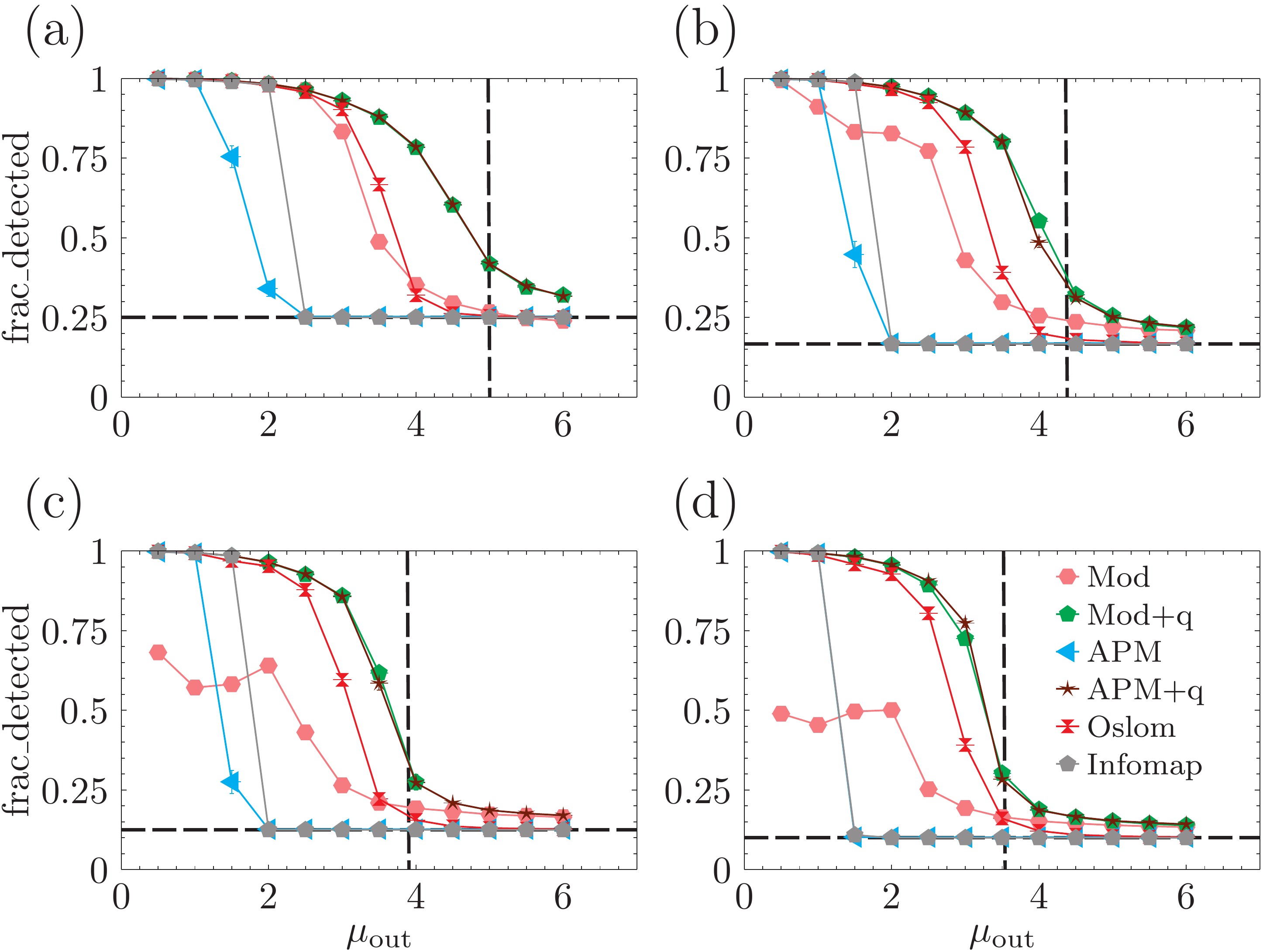}
\caption{(Color online) Fraction of correctly detected nodes for the
  planted partition model with $n=200$ nodes per cluster, average internal degree
  $\mu_{in}=10$ and four values for the number of clusters: (a) $q=4$,
  (b) $q=6$, (c) $q=8$ and (d) $q=10$. Symbols indicate
  the performance curves of different popular methods of community
  detection. For modularity optimization (Mod) and the Absolute Potts
  Model (APM) we show two curves, referring to the results of the
  method in the absence of any information on the number of clusters
  of the planted partition, and when such information is fed into the
  model as initial input. In both cases, knowing the number of
  clusters beforehand leads to a much better performance.  The
  detectability limit is shown as a vertical line which varies as a
  function of $q$, according to Eq. 1. The horizontal dashed line
  indicates the fraction $1/q$ of nodes correctly detected by a random
  partition of the graph in $q$ equal-sized clusters.}
\label{fig1B}
\end{center}
\end{figure}
For Mod and APM we performed two types of runs, one without any indication on
the number of clusters $q$ of the planted partition, the other by
constraining the optimization procedure to partitions with $q$
clusters (Mod+q and APM+q). Fig.~\ref{fig1} shows that in general the performance of all methods
is quite far from optimal. However, the limit performance (with slight
variations due to finite size effects) is attained
when the number of clusters $q$ of the planted partition is known. In particular, we remark that modularity
maximization, in the absence of any information on the number of
clusters, does a rather poor job. The limit curve and in general the
whole analysis by Nadakuditi and Newman assumes that one knows the
number of clusters beforehand, which is a very strong (and generally
invalid) hypothesis.
Furthermore, we observe that once the number of clusters is fixed to
the correct value (in Mod+q and APM+q), both methods return the same
results.  Practically speaking, the only thing which a method can then
do is move nodes to the cluster to which it has the most edges.  The
result is identical performance for any constant-$q$ method in this
simple symmetric case.  In Fig.~\ref{fig1B}, we see that knowledge of
$q$ remains important as the number of clusters increases.

\section{Inferring the number of clusters from spectra of graph matrices}
\label{sec3}

\begin{figure}[h]
\begin{center}
\includegraphics[width=\columnwidth]{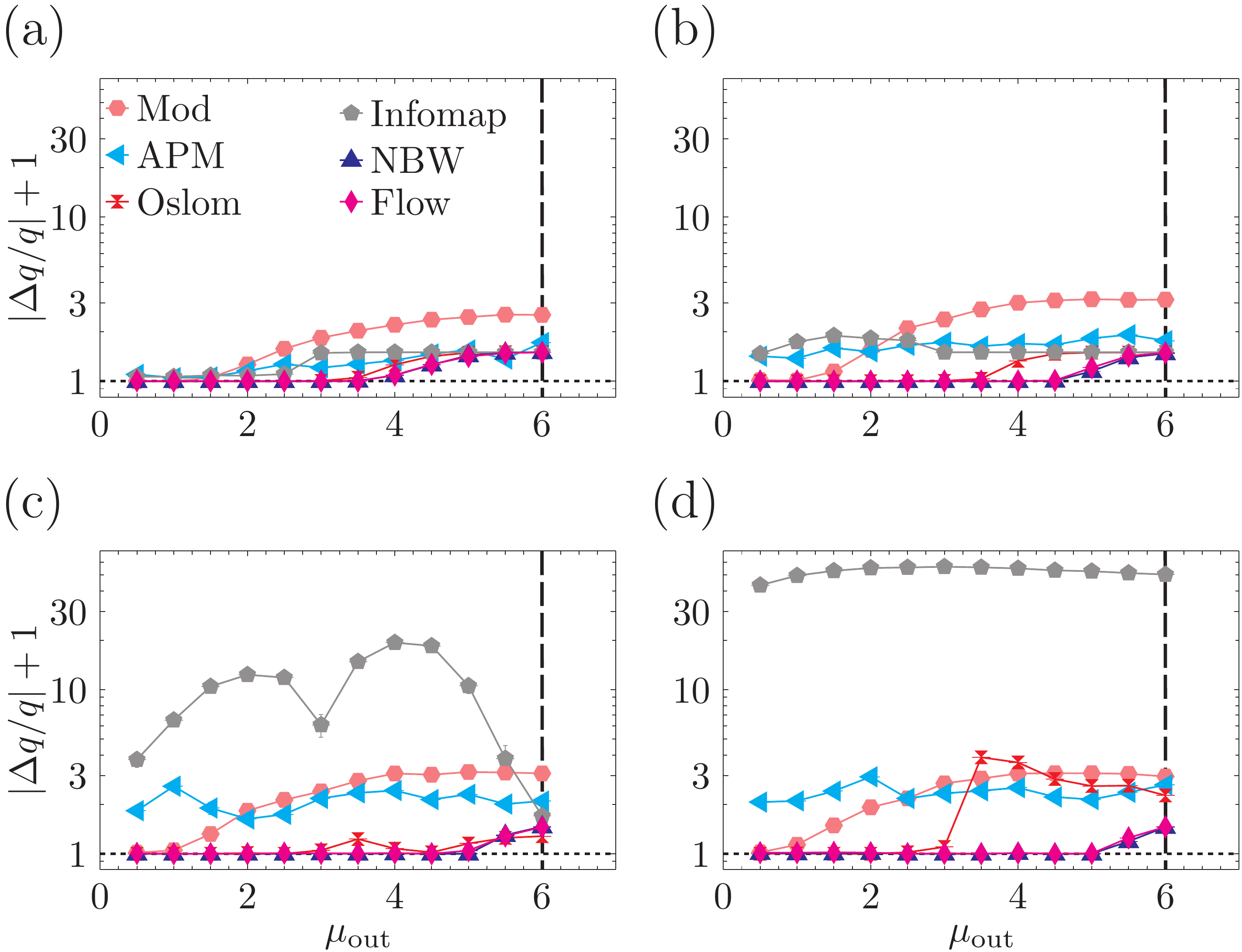}
\caption{(Color online) Accuracy of the prediction of the number of clusters by various methods as a
  function of the average external degree $\mu_{out}$, for the model
  graphs of Fig.~\ref{fig1}.  All graphs have $q=2$ clusters, average
  internal degree $\mu_{in}=10$, and the size varies: (a) $n=50$,
  (b) $n=200$, (c) $n=500$, and (d) $n=1000$.
  Appreciable deviations are found already at fairly
low values of $\mu_{out}$, when the planted partition is
pronounced. The predictions from the spectra of the non-backtracking
and flow matrices are more accurate and the results are better the
larger the graph size.  In the limit of infinite graphs it has been
conjectured that the number
of clusters inferred through the non-backtracking matrix matches the
right one all the way to the detectability
limit~\cite{krzakala13}.  The detectability limit is indicated as a
vertical dashed line.}
\label{fig2}
\end{center}
\end{figure}

From Section II we conclude that knowing the number of clusters beforehand could push
(some) methods up to the best attainable performance. Indeed, 
methods mostly fail in that they generally
find a partition with a different number of clusters than the planted
one. This is shown in Fig.~\ref{fig2}, where we show the average number of
clusters detected by the methods we used on the graphs of
Fig.~\ref{fig1}. For illustration purposes, we show the variable $|\Delta
q/q|+1$, where $\Delta q$ is the difference between the number
$q_{det}$ of detected clusters and the actual number $q$ (here
$q=2$). If $q_{det}=q$,  $\Delta q=q_{det}-q=0$ and $|\Delta
q/q|+1$=1. Naturally, Mod+q and APM+q yield the correct number of
clusters by default, for any $\mu_{out}$, since $q$ is given as input.
However, if the number of clusters is to be inferred by the method, it
may differ from the actual $q$ at $\mu_{out}$ far below the detectability limit. In
fact, for the largest graph sizes, deviations are large for very small
values of $\mu_{out}$, when the clusters of the planted partition are
connected by very few links and should be easily distinguishable. In particular, we notice that Infomap,
which is known to have an excellent performance on the more realistic
LFR benchmark~\cite{lancichinetti09c} breaks the two clusters in
many pieces, so it is not a valuable predictor of $q$ for this graph model.

On the other hand, it has been recently shown that 
the correct number of clusters in the planted partition model can be
correctly inferred from the spectra of two matrices: the
non-backtracking matrix $\bf B$ ~\cite{krzakala13} and the flow matrix
$\bf F$ ~\cite{newman13}.
Both are $2m\times 2m$ matrices, where $m$ is the number of links of
the network. Each link is considered in both directions, yielding $2m$
directed links and indicated with the notation $i\rightarrow j$,
meaning that the link goes from node $i$ to node $j$. 
Their elements read
\begin{equation}
B_{i\rightarrow j, k\rightarrow l}=\delta_{il}(1-\delta_{jk})
\label{eq2}
\end{equation}
and
\begin{equation}
F_{i\rightarrow j, k\rightarrow l}=\frac{\delta_{il}(1-\delta_{jk})}{d_i-1}.
\label{eq3}
\end{equation}
In Eq.~\ref{eq3} $d_i$ indicates the degree of node $i$. So the
elements of $\bf F$ are basically the elements of $\bf B$, normalized
by node degrees. This is done to account for the heterogeneous degree
distributions observed in most real networks. Both matrices have non-zero elements only for each
pair of links forming a directed path from the first node of
one link to the second of the other link. To do that, links have to
be incident at one node. As a matter of fact, the non-backtracking
matrix $\bf B$ is just the adjacency matrix of the (directed) links of
the graph.

A remarkable property of both matrices is that most eigenvalues, which
are generally complex, are enclosed by a circle centered at the
origin, and that the number of
eigenvalues lying outside of the circle is a good proxy of the number
of communities of the network. For $\bf B$ the circle's radius is
given by the square root $\sqrt{c}$ of the leading eigenvalue $c$; for
$\bf F$ it equals $\sqrt{\langle d/(d-1)\rangle/\langle d\rangle}$,
which is never greater than $1$. In Refs.~\cite{krzakala13} and
\cite{newman13} it was shown that the eigenvectors can be used for the
detection of the communities, by turning the nodes into points
in a $q-1$-dimensional Euclidean space ($q$ being the number of
off-circle eigenvalues, i.e. of clusters), and grouping them with
partitional clustering techniques, like K-means~\cite{macqueen67}.

In Fig.~\ref{fig2} we added the predictions of the number of clusters obtained
by both matrices (labeled by NBW and Flow). The exact techniques used
to compute the spectra is listed in Appendix~\ref{sec:methods}.
We find that the prediction is much more accurate than
those of all community detection methods we used, and it becomes more
precise, the larger the graph size, as expected. There is no
difference between the two curves, because the degree of all nodes are
essentially identical in the planted partition model, so the two
matrices are basically proportional to one another and their spectra
approximately coincide.

\begin{figure}[h]
\begin{center}
\includegraphics[width=\columnwidth]{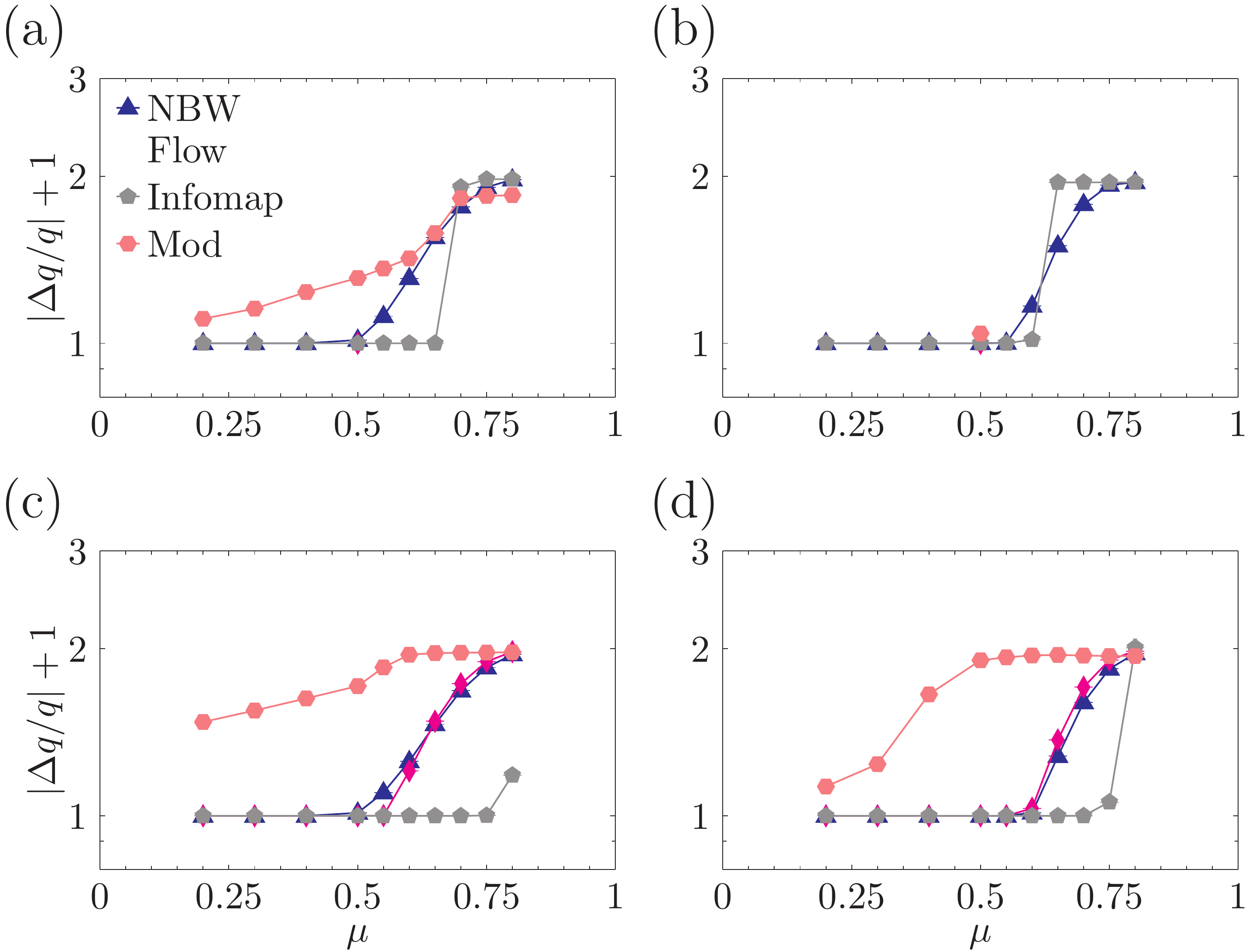}
\caption{(Color online) Accuracy of the prediction of the number of clusters by various methods as a
  function of the mixing parameter $\mu$ for LFR benchmark graphs of
  different sizes ((a,b) $1000$ and (c,d) $5000$ nodes) and cluster size ranges
  ((a,c) $[10:50]$ and (b,d) $[20:100]$).
  Infomap is the best predictor on these
  graphs, though the accuracy of the spectral methods seems to improve
if the system gets larger.}
\label{fig3}
\end{center}
\end{figure}

To see whether one can rely on the prediction of the number
of clusters from the spectra of $\bf B$ and $\bf F$, one should tackle
more complex network models with communities. The classic version of
the planted partition model, used here as well as in
Refs.\cite{krzakala13,newman13}, is too homogeneous to resemble any
real network. Nodes have approximately the same degree and communities
have exactly the same size. The LFR benchmark
graph~\cite{lancichinetti08} is an extension of the planted partition
model, where degrees and cluster sizes are distributed as power
laws, as in many real systems. We want to see whether the number of
clusters predicted by the
non-backtracking and the flow matrices is as reliable on the LFR
benchmark as it is on the classic model. This is shown in Fig.~\ref{fig3}.
The panels correspond to two different network sizes,
$1000$ and $5000$ nodes, and two different ranges for the cluster
sizes, S=small (communities comprise 10 to 50 nodes) and B = big
(communities comprise 20 to 100 nodes). The mixing parameter $\mu$ is the ratio of the external
degree of each node (with respect to the cluster it belongs to) by
the total degree of the node, so it varies from $0$ to $1$: values
near zero correspond to well-separated
clusters, whereas values near $1$ indicate a system with very mixed
communities (hence hard to identify).  The other parameters of the
LFR benchmark are: average degree = $20$, maximum degree = $50$,
degree distribution exponent = -2, cluster size distribution exponent
= -1. The average number of clusters is $\langle
q\rangle=40.334$ (1000, S),  $\langle
q\rangle=19.901$ (1000, B),  $\langle
q\rangle=203.239$ (5000, S),  $\langle
q\rangle=100.879$ (5000, B).  
These parameters are the same used for the comparative analysis
of community detection algorithms of Ref.~\cite{lancichinetti09c}. 
Infomap was the best-performing method on those graphs, and this is
reflected in Fig.~\ref{fig3}, as it gives the best prediction of the number of
clusters. Here the two matrices yield
close but distinct results, due to the inhomogeneity of the degree
distribution, but Infomap is clearly superior. We also remark that
modularity optimization (Mod) fails to guess $q$ also for very low
values of $\mu$, due to the resolution limit~\cite{fortunato07}.  So it seems that we do not
gain that much by using $\bf B$ and $\bf F$. 

\begin{figure}[h]
\begin{center}
\includegraphics[width=\columnwidth]{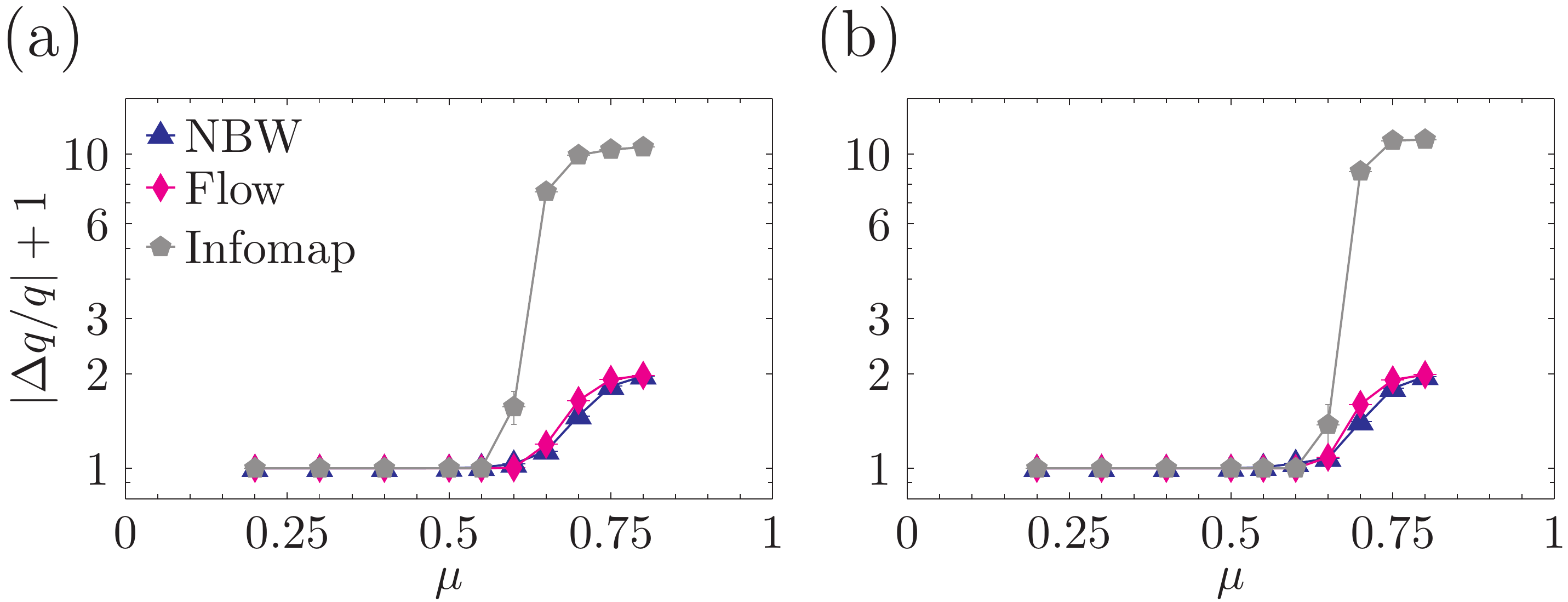}
\caption{(Color online) Same as Fig.~\ref{fig3}, but on LFR benchmark graphs
  with (a) $N=10000$ (50 averaging runs per point) and (b) $N=20000$ nodes
  (20 averaging runs per point). The range of community sizes
is $[10:1000]$. On these graphs the non-backtracking and flow
matrices, which give very similar results, outperform Infomap.}
\label{fig4}
\end{center}
\end{figure}

However, it may be that
the situation improves on larger systems. Therefore, we repeated the
procedure on two other sets of LFR graphs, with $10000$ and $20000$
nodes, respectively. We extended the range of cluster sizes to the
interval $[10: 1000]$, so that it spans two orders of magnitude and
there is a big difference between the smallest and the largest
community. All other graph parameters are the same as above.
Here the average number of clusters is $\langle
q\rangle=46.60$ (10000),  $\langle
q\rangle=95.66$ (20000).
In Fig.~\ref{fig4} we show that in these larger and
more heterogeneous graphs both the non-backtracking and the flow
matrices give a better prediction of $q$ than Infomap. 

\section{A two-step approach}
\label{sec4}

The results of Section III suggest that one might considerably improve
the performance of community detection algorithms with a two-step
procedure, in that one first infers the number of clusters through the
spectrum of the non-backtracking or the flow matrix and then
runs the algorithm on the space of partitions with that given number
of clusters. Even if the inference of the correct number of clusters
is not $100\%$ reliable, especially if the system is not too large,
one might still get much closer to the true partition than by running
the method without any information on the number of clusters.

To show that, we designed such two-step procedure for modularity
optimization.
The number of clusters is deduced through the flow matrix. Then,
modularity is optimized via a greedy procedure, which exploits the
idea of the Louvain algorithm~\cite{blondel08}. The latter works in a
hierarchical manner. First, each node is its own cluster. Then nodes are put in the clusters of their
neghbors such to yield the largest increase of modularity. This gives
the first hierarchical level. Then such
groups are turned into super-nodes and the procedure is repeated,
which means that the clusters of the first hierarchical level are combined into larger groups, yielding
the second hierarchical level, and so on,
until one reaches the partition with largest modularity. Naturally,
the number of clusters decreases when one moves from one
level/partition to the next. We stop at the level $L$ such that the
number of clusters $q_L$ is larger 
than the target number $q$ but the number of clusters $q_{L+1}$ of the level $L+1$
is smaller than $q$. If by any chance there is a level $L$ such that
$q_L=q$ we move to the refinement step described below.
If there is no level such that $q_L \geq q$, the algorithm is
restarted with a different seed until such a partition is found.
For the time
being, we assume to be in the more frequent case in which we start
with a partition with number of clusters larger than $q$.

We then perform the
following clean-up procedure in order to correct the number of
communities, decreasing it to the correct value $q$:
\begin{enumerate}
\item{Of all pairs of communities, choose the two which, when merged,
   give the greatest increase (least decrease) in the network
   modularity and merge them.}
\item{Repeat (1) until the number of communities equals $q$.}
\item{Iterate through all nodes in random order.  For each node,
   move it to the community which most increases modularity,
   or leave it in its present community if modularity can not be
   increased.}
\item{Repeat (3) until there are no further changes to be made.}
\end{enumerate}
\begin{figure}[h]
\begin{center}
\includegraphics[width=\columnwidth]{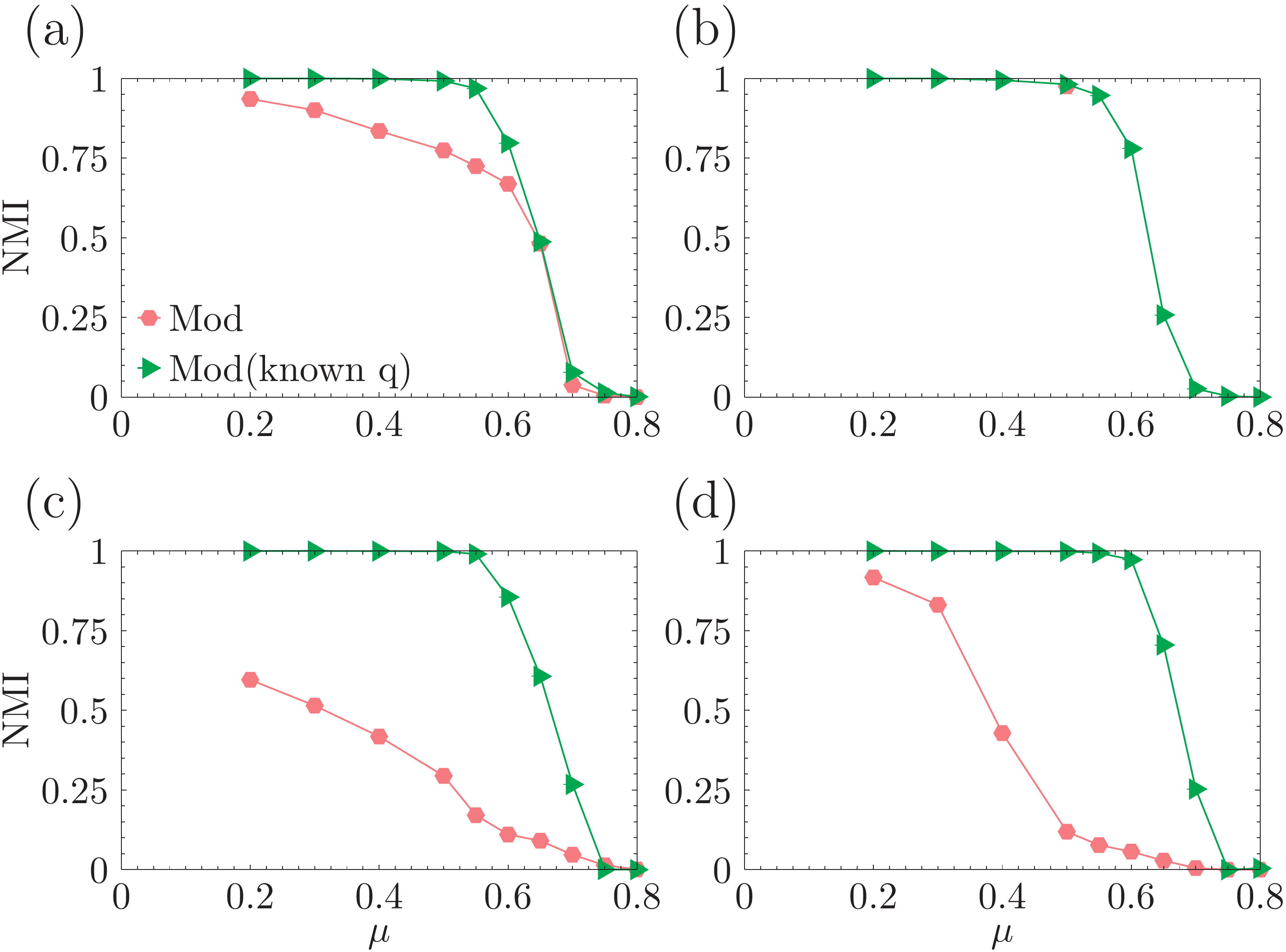}
\caption{(Color online) Modularity optimization on the LFR
  benchmark. The graphs are generated using the same parameters as for
  those in Fig.~\ref{fig3} for each (a-d), respectively. Hexagons indicate modularity optimization via
  simulated annealing, without any constraint on the number of clusters. The green
  triangles show the performance of our two-step procedure, where the
  number of clusters $q$ is inferred by the flow matrix and 
  modularity is optimized with a greedy procedure over the set of
  partitions with $q$ clusters. The improvement of the performance is
  manifest, especially on the larger graphs.}
\label{fig5}
\end{center}
\end{figure}

The last two steps are a refinement procedure aiming at further
improving the modularity without changing the number of clusters, and
are identical to the steps of our initial greedy procedure.
In Fig.~\ref{fig5} we show the performance of this procedure on LFR
benchmark graphs with the same parameters as the ones in Fig.~\ref{fig3} and
those used in the comparative analysis of
Ref.~\cite{lancichinetti09c}. As a reference, we report the
performance curve of the exhaustive maximization of modularity via
simulated annealing. We used simulated annealing because we wanted
to get a very good estimate of the actual modularity maximum.  
Here the performance is expressed by the
Normalized Mutual Information (NMI)~\cite{danon05}. We used the
extended version of the NMI proposed by Lancichinetti et
al.~\cite{lancichinetti09}, which can also compute the similarity of
covers, i.e. of partitions of the network into overlapping communities. 
This version has been used consistently throughout the comparative
analysis of Ref.~\cite{lancichinetti09c}. The two-step procedure
introduced here outperforms the unconstrained exhaustive optimization
of modularity, with the performance boost increasing for
larger graphs. In Table~\ref{flowtimes} we show that the two-step
algorithm has comparable complexity as simulated annealing.

\begin{table}[h]
  \centering
  \begin{tabular}{|c|c|c|c|}\hline
    N     & Spectral & Mod.          & Infomap \\ \hline
    1000  & $5 \pm 1$     & $30  \pm 10$  &$1 \pm 0.5$     \\\hline
    5000  & $250 \pm 10$  & $250 \pm 25$  &$3 \pm 1$      \\\hline
    10000 & $1200\pm 100$ & $1000\pm100$  &$8 \pm 1$    \\\hline
    20000 & $4700\pm 400$ & $4500\pm500$  &$18\pm 1$\\\hline
  \end{tabular}
  \caption{Comparison of the time complexity of several algorithms
    used in this work. Time is indicated in seconds. We took four
    different values for the graph size $N$ ($1000$,
    $5000$, $10000$ and $20000$), all other parameters of the LFR
    benchmarks are the same
    used for Fig.~4.  The actual
    computation time depends on the system parameters, not only on the
    size $N$.
    \textit{Spectral}: Time to calculate the largest $q$ eigenvalues
    of the flow matrix.  This time is far larger than the remaining
    constrained modularity optimization routine, so it is
    representative of the complexity of the full calculation. Computations were performed using
    \texttt{ARPACK}, which uses the implicitly restarted Arnoldi method
    for sparse matrix iterative eigenvalue computation. The complexity scales
    approximately as $t/s \sim N^{2.3}$. 
    \textit{Mod.}: Time to maximize modularity using simulated annealing with
    the parameters given in Sec.~\ref{sec:methods}.
    \textit{Infomap}: Time to run the Infomap algorithm.
    All computations were performed on an Intel(R) Core(TM)2
    Quad CPU (2.83 GHz) processor.}
  \label{flowtimes}
\end{table}

\section{Summary}
\label{sec5}

We have seen that the number of clusters is an important input for
community detection algorithms. If it is known beforehand, one can
push up the performance of methods. We have shown that the spectra of
the non-backtracking and flow matrices allow a reliable estimate
of the number of clusters, both on the classic planted partition model
and on the more realistic LFR benchmark, if the networks are not too
small. Therefore, a two-step procedure, when one first derives the
number of clusters and then performs a constrained run of the method
one wishes to use, might lead to much better results. We have shown
this for modularity optimization, whose unconstrained version
preferentially leads to clusters of a given scale, which may or may
not have anything to do with the actual scale of the clusters of the
system at study. Instead, once the number of clusters is given, and
the optimization is constrained on the set of partitions with that
number of clusters, the method gives much better results. 

In general, since the prediction of the number of clusters might not
be correct, one could constrain the method on the set of partitions
with number of clusters in a small range centered at the predicted
value. This way, even if the actual value is missed but close, it can
still be recovered by the procedure. It would be already valuable to
``push'' a method to the interesting range of values, instead of
letting it free to be attracted somewhere else by intrinsic biases, as
it happens for modularity.

On the practical side, computing the eigenvalues of the
non-backtracking or the flow matrix is lengthy. Both are 
$2m\times 2m$ matrices. The adjacency
matrix $\bf A$ has $N\times N$ elements, so $\bf B$ and $\bf F$ are
larger by a factor of
$\langle d\rangle^2$, where $\langle d\rangle$ is the average
degree of the network. Krzakala et al. have shown that the spectrum of
$\bf B$ can be computed by working on a $2N\times 2N$ matrix~\cite{krzakala13}, 
which reduces the complexity of the calculation by the factor $\langle
d\rangle^2$. Still, an approximate but reliable computation of the spectra requires a time which
scales approximately quadratically with the network size, by using
standard software libraries (see Appendix). This is because the time
needed to compute a single eigenvector/eigenvalue is linear in the
graph size, but the number of clusters $q$ is proportional 
to $N$, so the complexity of the calculation of $q$
eigenvectors/eigenvalues is approximately quadratic. Consequently, the problem is intractable for
graphs with number of links of the order of millions or higher.
This is the real bottleneck of the calculation. On the other hand,
many systems of interest remain within reach. Besides, one could find
other good predictors of the number of clusters which do not require
lengthy calculations. In the example we have seen of the LFR
benchmarks, Infomap is very good at guessing the right number of
clusters and it runs much faster than the computation of the spectra
of $\bf B$ and $\bf F$. However, we have seen that it gives poor
results on the classic planted partition model. Even if the latter is
not a good model of real networks with community structure, it means
that we should extensively test the reliability of a method before
using it in applications. This manuscript shows that it might be a
very good investment.

\section{Acknowledgements}
\label{sec6}

We thank Paolo De Los Rios, Lucio Floretta, Chris Moore, Lenka
Zdeborova and Pan Zhang for useful exchanges.
R. K. D. and S. F. gratefully acknowledge MULTIPLEX, grant number
317532 of the European Commission.

\appendix

\section{Methods}
\label{sec:methods}

In this section, we describe the community detection methods and used
in this work and their respective parameters.

\subsection{Modularity}

Network modularity is measure which rates the quality  of a network
partition, with higher modularity presumably indicating better
communities.  Modularity compares the actual number of
intra-community links to the expected number of links in a random
graph with the same degree distribution.  The details of modularity
optimization have been extensively described in other work, for
example~\cite{newman06}.  We adopt a simulated annealing approach in the
spirit of~\cite{guimera05}.  The Hamiltonian formulation of modularity
reads
\begin{equation}
  \mathcal{H}(\mathbf{\sigma}) =
  - \frac{1}{2m} \sum_{i < j}
  \left( A_{ij} - \frac{k_i k_j}{2m} \right)
  \delta( \sigma_i, \sigma_j )\,,
\label{eqa1}
\end{equation}
with $A_{ij}=1$ if there is an edge between nodes $i$ and $j$ and zero
otherwise, $k_i$, $k_j$ the degrees of nodes $i$ and $j$
respectively, $m$ the total number of (non-directed)
edges of the graph, $\sigma_i$, $\sigma_j$ the respective
community of nodes $i$ and $j$, and $\delta(i, j)$ the Kronecker delta
function.  A variety of trial moves are applied, and accepted
according to the Metropolis criteria~\cite{landau05}:
\begin{itemize}
\item \textit{Shift} moves: A node may be shifted from one community
  to another (possibly empty) community.  This may possibly result in
  the number of communities decreasing or increasing if the
  node is moved to a new community or was previously in a singleton
  community.
\item \textit{Merge} moves: Two randomly selected
  communities are selected and merged.
\item \textit{Split} moves: A community is
  chosen at random and split. When splitting, each node has an
  equal and independent
  probability of $1/2$ to join the new community being formed by the
  split.  A more sophisticated
  method could run community detection within the community in order
  to determine the optimal community split.
\end{itemize}

\textit{Mod}: For regular modularity calculations, first all nodes are
assigned to singleton communities (every node in its own distinct
community), and the Hamiltonian/energy is calculated from
Eq.~\ref{eqa1}.  Then, for each round, we do one trial move of each
node into a different random community ($N$ moves).  Then, there is
one merge move between two random communities, and one split move of a
random community.  After every round, the network modularity is
calculated and configuration is stored if the current partition is
better than the previously optimal partition.  We choose an initial
temperature $T_0=1/10$, and after each round, temperature is updated
via $T_\mathrm{new}=.999 T_\mathrm{old}$.  Trials continue until
energy does not decrease by more than $10^{-10}$ for 1000 rounds.
This entire procedure is repeated 5 times for each graph, and the
lowest energy result of all trials is selected as the final partition.
We achieved 100 averages for all points with these parameters.

\textit{Mod+q}: This procedure, used in Figs.~\ref{fig1}-\ref{fig2}, uses
simulated annealing with an additional term and different parameters
to fix the number of
clusters to $q_0$, the true number of communities.  The
Hamiltonian (note different normalization) is
\begin{equation}
  \mathcal{H}(\mathbf{\sigma}) =
  - \sum_{i < j}
  \left( A_{ij} - \frac{k_i k_j}{2m} \right)
  \delta( \sigma_i, \sigma_j ) + E_q\,,
\label{eqa2}
\end{equation}
with
\begin{equation}
  E_q = \epsilon\left( q-q_0 \right)^2
\end{equation}
being the term to fix $q$.  We choose $\epsilon=1$.  Thus, as
temperature decreases, the number of clusters becomes energetically
fixed at the true number.  The initial configuration is set with all
nodes divided randomly among $q_0$ clusters, though $q$ may fluctuate
during the minimization.  Each round consists of $N$ moves, with
$p_\mathrm{shift}=.95$, 
$p_\mathrm{merge}=.025$, and $p_\mathrm{split}=.025$ for each move.
We choose an initial temperature $T_0=100$, and after each round,
temperature is decreased by a factor
$T_\mathrm{new}=T_\mathrm{old}/1.001$. Trial rounds continue, until
there are no more changes accepted for 10 rounds.  This procedure is
repeated 10 times per graph, with the lowest final energy being taken
as the final partition.  We achieved 100 averaging runs for all points
with these parameters.

\subsection{Absolute Potts Model (APM)}

The APM detects communities via minimization of a Hamiltonian without
a null model,
\begin{equation}
  \mathcal{H}(\mathbf{\sigma}) =
  - \sum_{i < j}
  \left( A_{ij} - \gamma B_{ij} \right)
  \delta( \sigma_i, \sigma_j ),
\end{equation}
where $B_{ij}=1-A_{ij}$ is the inverse adjacency matrix.
In addition, $\gamma$ is a resolution parameter which can be
varied across a range of values in order to determine the optimal
community size.  The optimal $\gamma$ is inferred by the
self-similarity of multiple greedy trial minimizations performed on
the same graph.  We use the classic normalized mutual
information~\cite{danon05} to select the optimal $\gamma$, and in
general follow the
procedure of Ref.~\cite{ronhovde09}.  We achieved 100 averaging runs
for all points with these parameters.

\textit{APM+q}: For constant-$q$ APM runs, we do not use the
multi-resolution procedure of the previous paragraph.  Instead, we fix
$\gamma=10$ and use the constant-$q$ simulated annealing procedure
described under \textit{Mod+q} above, but with the APM Hamiltonian
plus $E_q$ term.  We achieved 100 averaging runs for all points with
these parameters.

\subsection{Oslom}

The Order Statistics Local Optimization Method (OSLOM) method seeks
statistically significant clusters in networks.  We use the OSLOM code
available at \url{http://oslom.org/}.  All default parameters are
accepted, but in addition we enable the option to ensure that all
nodes are a member of at least one community (no homeless nodes,
the default in the latest released code),
which is sensible for the case where all nodes are contained in at
least one community.  The default of 10 trial optimizations of the
lowest hierarchical level is used, and the lowest hierarchical level
is selected as the resulting partition.  All Oslom data is averaged
over 100 graph realizations unless otherwise indicated.

\subsection{Infomap}

Infomap is a method based on the compression of a random walk on
graphs.  The source code can be downloaded at
\url{http://mapequation.org/}.  All default options are accepted, but
notably we restrict the algorithm to a single non-hierarchical
partition (\texttt{--two-level}) and request ten trial minimizations
for each graph.  All Infomap data is averaged over 100 graph
realizations unless otherwise indicated.

\subsection{Spectral methods}

For a description of the non-backtracking walk and flow matrices, see
Sec.~\ref{sec3}.  We use a home-build implementation of these methods,
as described in Refs.~\cite{krzakala13,newman13}.  Actual eigenvalue
computation is via the \texttt{scipy} sparse linear algebra
package \cite{scipy}, which is an interface to the \texttt{ARPACK}
library~\cite{lehoucq98}, which itself uses the implicitly restarted
Arnoldi method for computing $k$ eigenvalues~\cite{lehoucq96}.  In
order to infer that a graph has $q$ communities, we must compute at
least $q+1$ eigenvalues ($q$ eigenvalues outside of the circle, and
one inside of the circle to know that the calculation is finished).
All spectral data is averaged over 100 graph
realizations unless otherwise indicated.
The time taken to compute $q+1$ eigenvalues scales as $t/s \sim
N^{2.3}$, as $q$ is generally a function of $N$ (approximately linear).  
All spectral data is averaged over 100 graph
realizations unless otherwise indicated.
Some actual times to compute these eigenvalues are shown
in Table~\ref{tab:flowtimes}.

\begin{table}[b]
  \centering
  \begin{tabular}{c|c|c|c|c}
    N     & Flow EV calc. & Mod.          & Infomap & Mod. known $q$\\ \hline
    1000  & $5 \pm 1$     & $30  \pm 10$  &$1 \pm.5$& $2\pm1$       \\
    5000  & $250 \pm 10$  & $250 \pm 25$  &$3 \pm 1$& $15\pm5$      \\
    10000 & $1200\pm 100$ & $1000\pm100$  &$8 \pm 1$& $90\pm30$     \\
    20000 & $4700\pm 400$ & $4500\pm500$  &$18\pm 1$& $-$
  \end{tabular}
  \caption{Computation times (in seconds) of various methods on
    LFR graphs with parameters parameters community size range $[20:50]$,
    mixing parameter $\mu=0.5$, average degree $20$, maximum degree $50$,
    degree distribution exponent $-2$, and cluster size distribution
    exponent $-1$.
    \textit{N}: Graph size, number of nodes.
    \textit{Flow EV calc.}: Time to calculate $q$ eigenvalues
    using the Flow matrix.  Only time spent doing the eigenvalue problem
    is included in this time.  Computations were performed using
    \texttt{ARPACK}, which uses the implicitly restarted Arnoldi method
    for sparse matrix iterative eigenvalue computation.
    \textit{Mod.}: Time to run modularity simulated annealing with
    parameters given in Sec.~\ref{sec:methods}.
    \textit{Infomap}: Time to run Infomap algorithm.
    \textit{Mod. known $q$}: Time to run the algorithm of
    Sec.~\ref{sec4}, \textit{excluding} the time to calculate
    eigenvalues, which are indicated in the second column.
    \textit{All}: Computations were performed on an Intel(R) Core(TM)2
    Quad CPU (2.83 GHz) processor.  Most calculations have large
    variations depending on the exact graph studied.}
  \label{tab:flowtimes}
\end{table}

\end{document}